\documentstyle[epsfig]{mn}

\newif\ifAMStwofonts

\def\gtorder{\mathrel{\raise.3ex\hbox{$>$}\mkern-14mu
             \lower0.6ex\hbox{$\sim$}}}
\def\ltorder{\mathrel{\raise.3ex\hbox{$<$}\mkern-14mu
             \lower0.6ex\hbox{$\sim$}}}

\ifoldfss
  \ifCUPmtlplainloaded \else
    \NewTextAlphabet{textbfit} {cmbxti10} {}
    \NewTextAlphabet{textbfss} {cmssbx10} {}
    \NewMathAlphabet{mathbfit} {cmbxti10} {} 
    \NewMathAlphabet{mathbfss} {cmssbx10} {} 
  \fi
  \ifAMStwofonts
    \ifCUPmtlplainloaded \else
      \NewSymbolFont{upmath} {eurm10}
      \NewSymbolFont{AMSa} {msam10}
      \NewMathSymbol{\upi}     {0}{upmath}{19}
      \NewMathSymbol{\umu}     {0}{upmath}{16}
      \NewMathSymbol{\upartial}{0}{upmath}{40}
      \NewMathSymbol{\leqslant}{3}{AMSa}{36}
      \NewMathSymbol{\geqslant}{3}{AMSa}{3E}

    \fi
  \fi
\fi 

\ifnfssone
  \newmathalphabet{\mathit}
  \addtoversion{normal}{\mathit}{cmr}{m}{it}
  \addtoversion{bold}{\mathit}{cmr}{bx}{it}
  \newmathalphabet{\mathbfit} 
  \addtoversion{normal}{\mathbfit}{cmr}{bx}{it}
  \addtoversion{bold}{\mathbfit}{cmr}{bx}{it}
  \newmathalphabet{\mathbfss} 
  \addtoversion{normal}{\mathbfss}{cmss}{bx}{n}
  \addtoversion{bold}{\mathbfss}{cmss}{bx}{n}
  \ifAMStwofonts
    \ifCUPmtlplainloaded \else
      \UseAMStwoboldmath
      \makeatletter
      \new@mathgroup\upmath@group
      \define@mathgroup\mv@normal\upmath@group{eur}{m}{n}
      \define@mathgroup\mv@bold\upmath@group{eur}{b}{n}
      \edef\UPM{\hexnumber\upmath@group}
      \new@mathgroup\amsa@group
      \define@mathgroup\mv@normal\amsa@group{msa}{m}{n}
      \define@mathgroup\mv@bold\amsa@group{msa}{m}{n}
      \edef\AMSa{\hexnumber\amsa@group}
      \makeatother
      \mathchardef\upi="0\UPM19
      \mathchardef\umu="0\UPM16
      \mathchardef\upartial="0\UPM40
      \mathchardef\leqslant="3\AMSa36
      \mathchardef\geqslant="3\AMSa3E
    \fi
  \fi
\fi 

\ifnfsstwo
  \DeclareMathAlphabet{\mathbfit}{OT1}{cmr}{bx}{it}
  \SetMathAlphabet\mathbfit{bold}{OT1}{cmr}{bx}{it}
  \DeclareMathAlphabet{\mathbfss}{OT1}{cmss}{bx}{n}
  \SetMathAlphabet\mathbfss{bold}{OT1}{cmss}{bx}{n}
  \ifAMStwofonts
    \ifCUPmtlplainloaded \else
      \DeclareSymbolFont{UPM}{U}{eur}{m}{n}
      \SetSymbolFont{UPM}{bold}{U}{eur}{b}{n}
      \DeclareSymbolFont{AMSa}{U}{msa}{m}{n}
      \DeclareMathSymbol{\upi}{0}{UPM}{"19}
      \DeclareMathSymbol{\umu}{0}{UPM}{"16}
      \DeclareMathSymbol{\upartial}{0}{UPM}{"40}
      \DeclareMathSymbol{\leqslant}{3}{AMSa}{"36}
      \DeclareMathSymbol{\geqslant}{3}{AMSa}{"3E}
    \fi
  \fi
\fi 

\ifCUPmtlplainloaded \else
  \ifAMStwofonts \else 
    \def\upi{\pi}
    \def\umu{\mu}
    \def\upartial{\partial}
  \fi
\fi

\title[Supernova 2002ap - The First Month] {Supernova 2002ap - The First Month}

\date{Accepted - .
      Received - ;}

\author[A. Gal-Yam et al.]{Avishay Gal-Yam$^{1,2}$, 
Eran O. Ofek$^{1}$,
and Ohad Shemmer$^{1}$\\
$^{1}$ School of Physics \& Astronomy and Wise Observatory, Tel Aviv University, Tel Aviv 69978, Israel; avishay@wise.tau.ac.il \\
$^{2}$ Colton Fellow. \\}

\begin{document}

\maketitle


\begin{abstract}

Supernova (SN) 2002ap in M74 was discovered on January 29, 2002.
Being one of the nearest (10 Mpc) SN events in the last 
decades, and spectroscopically similar 
to the so-called ``hypernovae'' 1997ef and 1998bw, both possibly
associated with gamma-ray bursts (GRBs), it is of great
interest. Shortly after its discovery,
we launched an intensive photometric and spectroscopic monitoring campaign
of this event, and here we report the results of the first
month of observations. We use our $UBVRI$ photometry to
estimate the magnitudes at, and dates of, peak brightness. 
Our data suggest that this object reached its peak $B-$band luminosity on 
Feb. $7.1_{-1.3}^{+2}$ UT. Based on its similarity to SN 1998bw, we estimate 
the range of possible 
dates for a GRB that may have been associated with SN 2002ap. We find 
that it may include dates outside the time frame for which all available 
gamma-ray data have been intensively scanned, according to recent reports. 
The absolute magnitude at peak 
brightness of SN 2002ap (M$_{B}=-16.9$) shows that it was
significantly fainter than SN 1998bw, or normal
type-Ia SNe, but similar to SN 1997ef. 
Our spectroscopic observations confirm that SN 2002ap
is strikingly similar to SNe 1998bw and 1997ef. We briefly describe the 
spectral evolution of this object. To assist other observers 
and to stimulate theoretical models, we make our entire 
data set publicly available in digital form.\footnotemark[1]

\end{abstract}

\begin{keywords}
supernovae: individual: SN 2002ap -- gamma-rays: bursts.
\end{keywords}


\section{Introduction}

Supernova (SN) 2002ap in M74 was discovered by Y. Hirose and confirmed by 
R. Kushida and W. Li in images obtained on January 29 and 30, 2002 
(Nakano et al. 2002; UT dates are used throughout this {\it Letter}). 
The relative proximity of this event ($v_{r}=657$ km s$^{-1}$, $D\sim10$ 
Mpc, Tully 1988) and its ensuing brightness made it a promising target 
for intensive follow-up
studies. Three independent teams promptly obtained low resolution 
spectroscopic observations on Jan. 30 (Meikle et al. 2002), and Jan. 31
(Kinugasa et al. 2002, Gal-Yam \& Shemmer 2002). All groups noted the
spectral resemblance of this object to the so-called ``hypernovae'' 1998bw
and 1997ef, both of which may have been associated with gamma-ray
bursts (GRBs; Galama et al. 1998, Iwamoto et al. 2000). Using the well-sampled
spectral observations of SN 1998bw by Patat et al. (2001), 
Gal-Yam \& Shemmer estimated that SN 2002ap was discovered prior to peak
brightness. Being both nearby, of a rare type, and possibly connected
to GRBs, SN 2002ap became a focus of a multi-wavelength observational
effort. Follow up studies include detection in the radio 
(Berger, Kulkarni, \& Frail 2002), IR imaging (e.g. Mattila \& Meikle 2002) 
and spectroscopy (e.g., Motohara et al. 2002), UV and X-ray imaging 
(Rodriguez-Pascual et al. 2002), optical polarimetry (e.g., Wang et 
al. 2002) and high-resolution spectroscopy (e.g., Lauroesch et al. 2002).

We have undertaken an intensive observational
monitoring campaign of this SN with multi-color imaging and 
low-resolution optical spectroscopy. The backbone of our program consists of
target-of-opportunity observations obtained with the Wise Observatory 
1-m telescope as part of the queue program.
The resulting set of observations presents a coherent 
picture of the evolution of SN 2002ap in the optical band during the first 
month since its discovery. As this SN now approaches solar conjunction
and will not be observable during the next few months, we have analysed
the data set collected so far. In this {\it Letter} we briefly report 
and discuss our main results. All the data presented here are available
electronically via our website\footnotemark[1].

\footnotetext[1]{http://wise-obs.tau.ac.il/$\sim$avishay/local.html}

\section{Observations}

\subsection{Photometry}

$UBVRI$ photometry was obtained with the Wise Observatory 1-m 
telescope during nine nights, using the
Tektronics ($1024\times1024$ pixel)
and SITe ($2048\times4096$ pixel) CCD cameras. 
Time series of 60 s $V-$band images were
obtained on five occasions, in order to search for variability on 
short ($< 1$ hr) time scales. As no such effects were evident, these
data were averaged and used to improve the photometric sampling. 

Figure 1 shows a combined $BVR$ image of the field of SN 2002ap, produced from
the stacked images obtained during February 2002. As the SN is isolated
and contamination by underlying galactic light is constrained to lie below
$B=21.6$ mag (Smartt, Ramirez-Ruiz, \& Vreeswijk 2002), we used aperture photometry on the SN and nearby reference stars (see Fig. 1 and Table 1). 
The SN flux was measured and compared with that of nearby bright stars
using standard IRAF\footnotemark[2] routines. The photometric stability of each
reference star in each band was checked against the entire reference set, and stars 
with the lowest root-mean-square scatter values were used in the final calibration. $1-\sigma$ error estimates, including
\footnotetext[2]{IRAF (Image Reduction and Analysis Facility) is distributed
by the National Optical Astronomy Observatories, which are operated
by AURA, Inc., under cooperative agreement with the 
National Science Foundation.}
Poisson noise in the SN flux and the scatter in the flux of reference 
stars, were computed, and found to be below 0.025 mag, except for the 
$U-$band data, where some points have errors as large as 0.16 magnitudes.    
In Figure 2 we plot the resulting $UBVRI$ light curves of this object.  
The data provided by Henden (2002) were used for absolute calibration. 
We estimate the errors in the absolute zero 
point determination to be smaller than $2\%$. As a consistency check, we 
compare our photometry with the measurements of Riffeser, Goessl, \& Ries, 
(2002) and find excellent agreement. We therefore combine their early 
$UBR$ points with our data set. As both our own observations and those of
SN 1998bw around peak brightness (Galama et al. 1998) suggest that these
objects have smoothly varying light curves, we use cubic spline interpolation 
to get approximate $UBVRI$ curves. From these, we determine the dates of
peak brightness, and the SN magnitudes at peak in each band. 
Conservative margins on the peak dates are set by 
the latest point seen to be rising and the first to be declining. 
Assuming that the light curves of SN 2002ap had similar shapes around
peak to those of SN 1998bw, 
we can better constrain the dates of peak by checking by how much one can 
adjust the peak dates without creating pronounced ``wiggles'' or ``humps'', 
not seen in the light curves of Galama et al. (1998)
or in well sampled parts of our own data.
The calculated peak dates and their estimated errors, as well as the 
more conservative margins discussed above, are reported in
Table 2, along with the SN absolute magnitudes, calculated assuming the
distance modulus to M74 given by Tully (1988), $\mu=30.24$ for $H_{0}=65$ km s$^{-1}$ Mpc$^{-1}$.  
We correct for foreground Galactic extinction according to Schlegel,
Finkbeiner, \& Davis (1998), using the extinction curve of Cardelli, 
Clayton, \& Mathis (1989). Based on high resolution
spectroscopic observations, Klose, Guenther \& Woitas (2002) estimate that 
this SN suffered little intrinsic extinction ($A_{V}\sim0.025$ mag) 
in its host, so the absolute 
magnitudes we derive are probably a fair estimate of the
SN luminosity. If so, it is clearly evident that unlike SN 1998bw, 
SN 2002ap was not optically luminous, being $\sim2$ magnitudes fainter
in the $B-$band. The peak luminosity of SN 2002ap may be comparable to
that of SN 1998bw only if the distance to M74 is significantly 
underestimated. Apart from
the lower peak magnitudes, the general shape of the light curves is 
similar to those of SN 1998bw (Galama et al. 1998). As can be seen 
in Fig. 2 and Table 2, the light curves are wider in the redder bands, and 
the bluer the band, the earlier the date of the peak. The decline rates are 
also similar, with the SN becoming 
fainter by $\sim1.3$ mag in the $V-$band some 
20 days after peak.  

\begin{figure}
\centerline{\epsfxsize=85mm\epsfbox{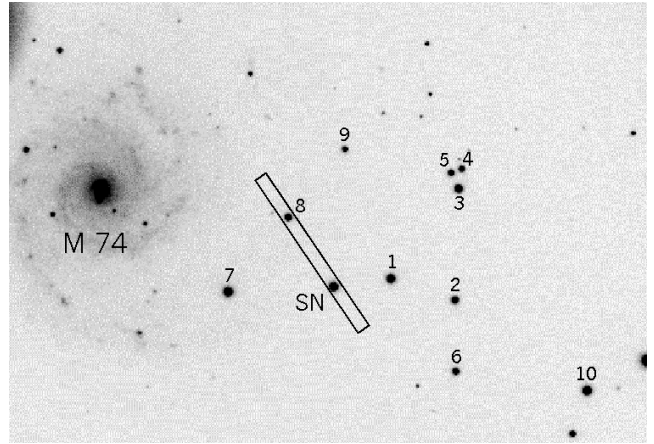}}
\caption{A $12'$ by $8.3'$ section of a combined $BVR$ image of the field of
SN 2002ap, obtained at the Wise Observatory. Nearby reference
stars are marked, as well as the schematic location of the spectroscopic
comparison star within the slit.}
\end{figure}

\begin{figure}
\centerline{\epsfxsize=85mm\epsfbox{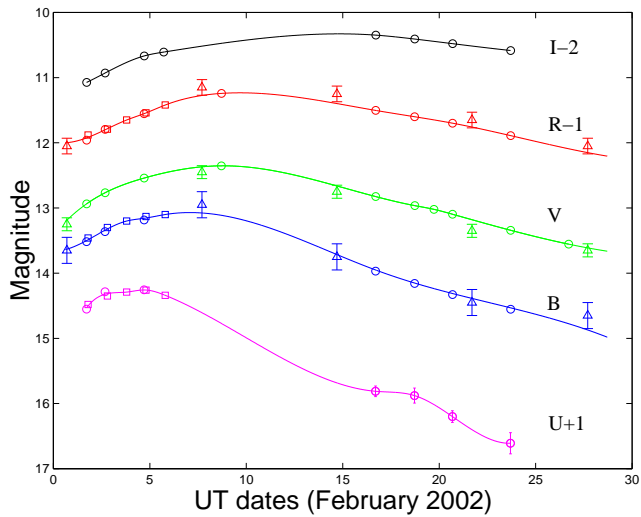}}
\caption{Multi-color light curves of SN 2002ap. Note the excellent
agreement between the Wise photometry (circles) and the
photometry of Riffeser, Goessl and Ries (2002; squares). For all the 
points lacking error bars, the estimated errors are smaller than the size 
of the mark. Photometric points calculated synthetically
from Wise spectroscopy (see Poznanski et al. 2002 for details)
are marked with triangles. Error bars reflect the combined 
uncertainties resulting from the flux calibration 
of the spectra and, in some cases, incomplete spectral coverage of the filter 
bandpasses. The solid lines are cubic-spline interpolations between
the photometric points.
For clarity, the $U$, $R$ and $I$ light curves have been shifted by
the amounts noted on the plot.}
\end{figure}

\begin{table}
\caption {Local calibrators in the field of SN 2002ap} \label {stars table}
\vspace{0.2cm}
\begin{centering}
\begin{minipage}{70mm}
\begin{scriptsize}
\begin{tabular}{lllccccc}
\hline
$\#$ & R.A. & Dec. & $U$ & $B$ & $V$ & $R$ & $I$ \\
\hline
1 & 01:36:19.53 & 15:45:21.8 & 14.10 & 13.84 & 13.06 & 12.61 & 12.14 \\
2 & 01:36:14.61 & 15:44:58.2 & 14.40 & 14.33 & 13.69 & 13.32 & 12.95 \\
3 & 01:36:14.32 & 15:47:01.5 & 14.30 & 14.01 & 13.22 & 12.77 & 12.34 \\
4 & 01:36:14.09 & 15:47:23.7 & 15.30 & 15.10 & 14.37 & 14.37 & 13.56 \\
5 & 01:36:14.90 & 15:47:19.4 & 15.69 & 15.22 & 14.38 & 13.89 & 13.45 \\
6 & 01:36:14.56 & 15:43:38.9 & 15.32 & 14.83 & 13.90 & 13.40 & 12.92 \\
7 & 01:36:32.05 & 15:45:07.8 & 15.27 & 14.11 & 12.85 & 12.07 & 11.35 \\
8 & 01:36:27.38 & 15:46:30.3 & 14.73 & 14.53 & 13.80 & 13.37 & 12.97 \\
9 & 01:36:23.04 & 15:47:45.5 & 15.17 & 15.18 & 14.61 & 14.26 & 13.90 \\
10 & 01:36:04.48 & 15:43:17.7 & 13.82 & 13.61 & 12.91 & 12.52 & 12.16 \\
\hline
\end{tabular}
\end{scriptsize}

Notes:\\ 
Astrometric solution based on 33 USNO-A2.0 (Monet et al. 1996) 
stars with final RMS scatter of $0.331''$ and $0.462''$ in R.A. 
(J2000) and Dec., respectively; Photometric $UBVRI$ data are 
from Henden (2002); Star $\# 8$ is the spectroscopic comparison star.
\end{minipage}
\end{centering}
\end{table}

\begin{table}
\caption {SN 2002ap peak dates and magnitudes} \label {mags table}
\vspace{0.2cm}
\begin{centering}
\begin{minipage}{70mm}
\begin{tabular}{cccccc}
\hline
 & $U$ & $B$ & $V$ & $R$ & $I$ \\
\hline
Peak date & $4.7^{+1}_{-1}$ & $7.1^{+2}_{-1.3}$ & $8.8^{+2}_{-2}$ & $9.7^{+2}_{-2}$ & $14.8^{+1}_{-4}$ \\
(Feb. 2002) & & & & & \\
Possible range & $_{-1.9}^{+1.1}$ & $_{-1.3}^{+7.6}$ & $_{-4}^{+5.9}$ & $_{-3.9}^{+7.9}$ & $_{-9}^{+3.9}$\\
(days) & & & & & \\
Apparent & 13.3 & 13.1 & 12.4 & 12.2 & 12.3 \\
magnitude & & & & & \\
Absolute & $-16.6$ & $-16.9$ & $-17.7$ & $-17.8$ & $-17.8$ \\
magnitude & & & & & \\
Adopted & 0.387 & 0.307 & 0.236 & 0.190 & 0.138 \\
extinction & & & & & \\
\hline
\end{tabular}

Note:\\ 
The given peak dates were estimated from cubic spline interpolations
and their errors estimated assuming that the light curves of SN 2002ap
were similar to those of SN 1998bw around peak (see text). 
The most conservative range of possible dates for the peak in each band 
was set to lie between the latest point seen to be rising and the
earliest seen to be on the decline.\\
\end{minipage}
\end{centering}
\end{table}

\subsection{Spectroscopy} 

Long-slit spectra of SN 2002ap were obtained with the Wise Observatory
1m telescope on five nights: 2002 January 31 and February 7, 14, 21 and 
27, using the Faint Object Spectrograph and Camera (FOSC; Brosch \& Goldberg
1994). We used a 10$''-$wide slit, and a 600 line mm$^{-1}$ grism,
resulting in a dispersion of 3.75~\AA/pix ($\sim$8~\AA\ resolution) and
a spectral range of $\sim$4000 -- 7800~\AA. Two exposures were obtained 
each night and used to reject cosmic ray hits.
Reduction of the bias-subtracted and flat-field corrected spectra was
carried out in the usual manner using IRAF\footnotemark[2].
A He-Ar arc spectrum was used for the wavelength solution.
In all but the first spectrum, the SN and a nearby bright star 
were observed simultaneously through the slit (see Fig. 1), providing an
intrinsic relative calibration during non-photometric conditions
(see, e.g., Kaspi et al. 2000).
Our photometry shows this star to be constant (to within photometric errors)
during the period of the observations. The wide ($10''$) slit ensured 
that the effects of atmospheric refraction at the non-parallactic angle were 
not important.  
Each spectrum of the SN was divided by the simultaneously-observed
spectrum of the comparison star. A spectrum of the comparison star obtained
on Feb. 7, under photometric conditions, was flux calibrated using the Wise
Observatory standard sensitivity function and extinction curve. These
do not change appreciably from night to night, and are routinely updated 
using spectrophotometric standard stars.
The SN spectra were calibrated to an absolute flux scale by multiplying each 
SN/star ratio by the flux calibrated spectrum of the comparison star. 
The January 31 spectrum (taken without the comparison star in the slit) 
was directly calibrated using the mean curves. The final fluxed
spectra were then compared to our photometry (Fig. 2), showing excellent
consistency. We estimate that the relative flux calibration of the last four
spectra is good to $2-3$ per cent, and the absolute flux
calibration has an uncertainty of $\sim$10 per cent, at most.

Figure 3 shows our entire spectroscopic data set. The first spectrum
is very blue and almost featureless, peaking around $5000$~\AA~with
secondary broad emission peaks around $4200$~\AA~and $6200$~\AA. This
spectral shape closely resembles early time spectra of SN 1997ef
(e.g. Garnavich et al. 1997)
and 1998bw (e.g., Patat et al. 2001, see also Fig. 4, top panel) leading to the 
prompt spectral identification of this object ($\S~1$). Figure 4 
demonstrates that this spectral resemblance continues throughout the epoch of
peak brightness (unless otherwise noted, from here on we refer to our 
best estimate $B-$band peak date, Feb. 7.1 UT, as the time of peak brightness), 
and up until the date of our last spectrum ($\sim20$ days 
after maximum).  

The spectral evolution of SN 2002ap displays several
characteristic trends also identified in 1997ef (Matheson et al. 2001, 
Iwamoto et al. 2000) and 1998bw (Stathakis et al. 2000,
Patat et al. 2001, and references therein). These include
the reddening of the continuum, a redward shift and narrowing of 
prominent ``emission-like'' and ``absorption-like''\footnotemark[3] features, 
and decreasing values of the expansion velocities. 
\footnotetext[3]{Model synthetic spectra show that at early times,
the spectral shape of these objects is determined by the wavelength 
dependence of the absorption optical depth, so that features in the
spectrum cannot generally be attributed to a certain emission or
absorption line (e.g. Patat et al, 2001, and references within.)} 
Following Patat et al. (2001)
we have calculated expansion velocities from the Si II $\lambda 6355$ 
absorption, and find a decrease from $\sim 38000$ km s$^{-1}$ 
$7$ days prior to peak to $\sim 15000$ km s$^{-1}$ at peak, 
and to $\sim 6000$ km s$^{-1}$ 14 days later. This is
similar to the results obtained for 1998bw, and consistent with the
measurements of Motohara et al. (2002; $\sim 16000$ km s$^{-1}$ around peak
from $1.083 \mu$ He~I) and Filippenko \& Chornock (2002; $\sim 9000$
km s$^{-1}$ four days after peak from $7774$~\AA~O~I). 
The prominence of the $4600$~\AA~emission peak
relative to the one bluewards of $5000$~\AA~(both tentatively attributed to
Fe II blends, e.g., Patat et al. 2001 and references therein)
in spectra of SN 2002ap at and after peak resembles spectra of SN 1997ef but
differs somewhat from those of SN 1998bw. A notable feature in 
the spectra of SN 2002ap is a narrow, unresolved absorption, consistent with 
He II $\lambda4686$. This feature first emerges in the spectrum taken
near peak brightness and is stronger still seven days later. It does not appear
in later spectra of the object, nor in spectra of either SN 1997ef or 1998bw. 

\begin{figure}
\centerline{\epsfxsize=85mm\epsfbox{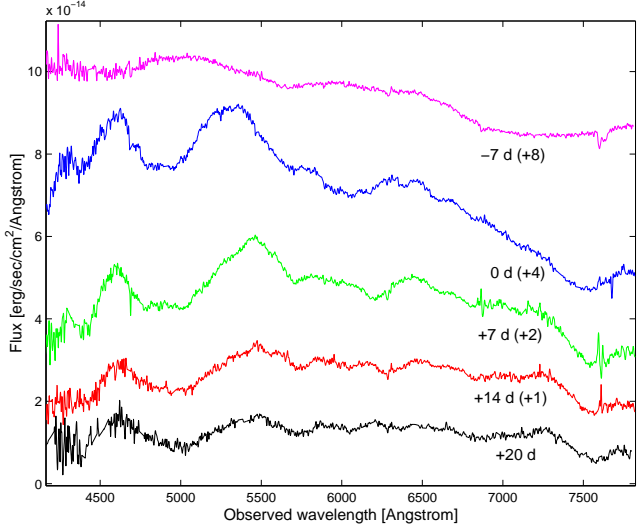}}
\caption{Time sequence of Wise Observatory spectra of SN 2002ap.
For each spectrum we note the SN age relative to our best estimate
$B-$band maximum, Feb. 7.1. For clarity, some of the spectra were
vertically shifted by the amounts noted in parenthesis, in units of 
$10^{-14}$ erg s$^{-1}$ cm$^{-2}$~\AA$^{-1}$. We have replaced short regions in
the spectra, strongly affected by cosmic-ray hits, with linear 
interpolations. Features around $6300$~\AA, 
$6900$~\AA~and $7600$~\AA, are residuals from 
imperfectly removed sky lines and telluric absorption.}
\end{figure}
 
\begin{figure}
\centerline{\epsfxsize=85mm\epsfbox{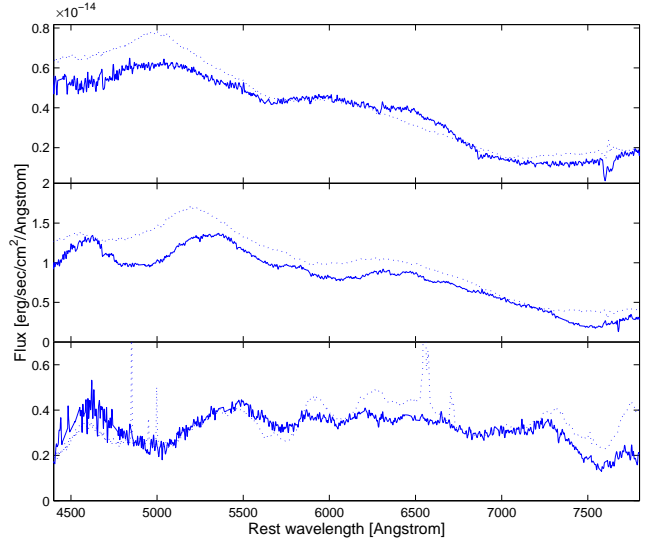}}
\caption{Comparison of Wise spectra of SN 2002ap (solid lines)
with contemporary spectra of SN 1998bw (dotted line)
at ages $7$ days before peak (top panel)
and at peak magnitude (middle panel), from Patat et al. 2001. 
The bottom panel demonstrates the resemblance of the spectrum of
SN 2002ap at age 20 days (solid line) to SN 1997ef
at age 40 days (dotted line; from Matheson et al. 2001; the superposed 
narrow lines are from the underlying galaxy;
earlier spectra of SN 1997ef were not available in digital form).}
\end{figure}

\section{Discussion and Conclusions} 

We have presented the results of our optical photometric and
spectroscopic monitoring of SN 2002ap, during the first month after its 
discovery. Our best estimate for the date of
$B-$band maximum is Feb. 7.1, 2002, and our derived peak magnitude 
of M$_{B}=-16.9$ is 
significantly lower than the one measured for the ``hypernova'' SN 1998bw.
However, except for the lower peak magnitude, the 
characteristics of the light curves, as well as the spectral evolution
of this object, are strikingly similar to those of SNe 1997ef and 1998bw.
Our work shows that, at least to some degree, one may use the well studied
properties of SN 1998bw as a tentative basis upon which to plan future
observations of SN 2002ap, after scaling the peak magnitude.  

In light of its similarity to SNe 1998bw and 1997ef, one of the 
intriguing questions about SN 2002ap is whether it was also associated
with a GRB. Using our most conservative 
estimate for the date of the $B-$band peak, 
Feb. $7.1_{-1.3}^{+7.6}$ days (Table 2), and assuming that the lag between a
hypothetical GRB and the time
of $B-$band maximum is similar to that measured for SN 1998bw (14.3 days,
Galama et al. 1998), we would expect the GRB trigger to have occurred around
Jan. $23.8_{-1.3}^{+7.6}$ days. 
However, if we use instead our estimated $U-$band peak 
date, Feb. $4.7_{-1.9}^{+1.1}$ days, which is 
best constrained by our photometry, 
along with the
appropriate $U-$band time lag from Galama et al. 1998 (13.7 days), the resulting
GRB trigger time is Jan. $22.0_{-1.9}^{+1.1}$ days. 
Hurley et al. (2002) found no candidate GRB that might be associated with
SN 2002ap in an intensive search of gamma-ray data from 
all available sources, starting Jan. 21. Our results suggest that 
the GRB trigger may have occurred outside the time frame searched. 
If the GRB-peak magnitude time lag for SN 2002ap was just one day longer
than the lag measured for SN 1998bw, the trigger is likely to have been
missed by Hurley et al. We conclude that in order to detect, or set a secure
upper limit to the fluence of a GRB associated with SN 2002ap, a search similar
to the one reported by Hurley et al. should be extended to include data
taken several days prior to January 21.   
  
Finally, there remains the question of the classification of SN 2002ap and its
look-alikes, SNe 1998bw and 1997ef (and possibly also SN 1998ey, 
Garnavich, Jha, \& Kirshner 1998, and the recent SN 2002bl,
Filippenko, Leonard, \& Moran 2002). Various authors have used various
types for these events, usually some combination of Ib, Ic, and 
``peculiar''. In 
view of the fact that these SNe are spectroscopically similar to each other,
but well
distinguished from prototypical SNe Ib (e.g. SN 1984L) or Ic (e.g., SN 1987M or
1994I), we suggest that they be
defined as a new SN sub-type, and provisionally designate them as type Id SNe. 
This should replace the often used nickname ``hypernovae'', initially 
used to describe the unusually energetic and luminous SN 1998bw
(Iwamoto et al. 1998). This term has since become somewhat misleading as it 
was later applied to events that were modeled as highly energetic 
explosions, but were not especially luminous, such as SN 1997ef 
(Iwamoto et al. 2000). Currently, all events that
are spectroscopically similar to SNe 1998bw and 1997ef, e.g.,
2002ap and 2002bl, are sometimes referred to as ``hypernovae'' for lack
of a better name, even though they are not known to be very luminous
or exceptionally energetic.     
Admittedly, these events are probably closer to type Ic SNe than to any other
sub-type, and the physical differences between them and other type Ic's are not
fully understood. However, the current SN classification scheme is purely 
observational and based mostly on optical spectroscopic properties 
(see Filippenko 1997, for a review). In our opinion, the difference between
the early-time spectra of the three well studied
``Id'' events and those of prototypical
Ib and Ic SNe are no less significant than the differences between other 
types of SNe. So much so that there was some question as to whether the 
first of these events, 1997ef, was in fact a SN, when judged by its earliest
spectra (Garnavich et al. 1997). 

\section*{Acknowledgments}

This work, and the queue observing program, would not have been possible 
without the support of the Wise Observatory staff, especially P. Ibbetson
and J. Dann. We thank N. Brosch, Y.-J. Choi, S. Kaspi and Y. Lipkin for their
help and support. Special thanks go to D. Maoz for many helpful discussions. 
A. V. Filippenko and F. Patat are thanked for making available
digital copies of their data on SNe 1997ef and 1998bw, and the anonymous
referee for useful comments. 
This work was supported by the Israel Science Foundation ---
the Jack Adler Foundation for Space Research, Grant 63/01-1.

\end{document}